\documentclass[english,journal=jacsat,manuscript=article,etalmode=truncate,maxauthors=0,layout=twocolumn]{achemso}
\setkeys{acs}{etalmode=truncate,maxauthors=0,articletitle=true,email=true}

\usepackage{amsmath}             % for equation typesetting
\usepackage{mathtools}             % for equation typesetting
\usepackage{amssymb}             % for equation typesetting
\usepackage{wasysym}             % for geometric shapes
\usepackage{color}               % for colored fonts
\usepackage{bm}
\usepackage{setspace}            % for 1.5 and double spacing
\usepackage{graphicx}            % main graphics package
\usepackage{diagbox}

\usepackage{caption}
\usepackage{subcaption}
\newcommand{\mycomment}[1]{}

\usepackage{wrapfig}             % allow text wrapping around figures
\usepackage[dvipsnames]{xcolor}  % for inserting colored text
\usepackage{array,booktabs,multirow}      % for formatting the journal option table at the end of this document
\usepackage{comment}
\usepackage[export]{adjustbox}
\usepackage{arydshln}
\usepackage{multirow, multicol}
\usepackage[ruled, vlined, linesnumbered]{algorithm2e}

\usepackage[font=footnotesize,labelfont=bf,labelsep=period,width=0.95\textwidth]{caption}   % format single-image captions and table titles
\usepackage[font=scriptsize,labelfont=bf]{subcaption}                     % format subfigure captions
\usepackage[capitalize]{cleveref}
\crefname{figure}{Figure}{Figures}
\Crefname{figure}{Figure}{Figures}
\crefname{table}{Tab.}{Tabs.}
\Crefname{table}{Table}{Tables}
\crefname{equation}{Eq.}{Eqs.}
\Crefname{equation}{Eq.}{Eqs.}
\crefname{section}{Sec.}{Secs.}
\Crefname{section}{Section}{Sections}

\newcommand{\bra}[1]{\left\langle #1 \right\vert}         % bra
\newcommand{\ket}[1]{\left\vert #1 \right\rangle}         % ket
\usepackage{xr}
\usepackage{etoolbox}% needed for the patch  <<<

\makeatletter
\renewcommand*{\acs@author@fnsymbol@symbol}[1]{% Use numbers instead of symbols, * is for email
    \ifcase #1 *\or
    1\or
    {\#}\or
    2\or
    3\or
    4\or
    5\or
    6\or
    7\or
    8\or
    9
    \fi
}
        
\renewcommand*\acs@contact@details{% addd * before  E-mail
    {\sffamily *\,Corresponding author; E-mail: \acs@email@list }%
    \acs@number@list
}           

\patchcmd{\acs@address@list@auxii}% superscript for numbers in affiliations
{\acs@author@fnsymbol{\acs@affil@marker@cnt}}
{\textsuperscript{\acs@author@fnsymbol{\acs@affil@marker@cnt}}}
{}{}

\patchcmd{\acs@address@list@auxii}% superscript for numbers in affiliations
{{\acs@author@fnsymbol{\acs@affil@marker@cnt}\@nameuse{@altaffil@\@roman\@tempcnta}\par}}
{{\textsuperscript{\acs@author@fnsymbol{\acs@affil@marker@cnt}}\@nameuse{@altaffil@\@roman\@tempcnta}\par}}
{}{}        

\makeatother    
\title[SPARSE DAS]{Numerically Exact Configuration Interaction at Quadrillion-Determinant Scale}

\author{Agam Shayit}%main developer
\affiliation[University of Washington]
{Department of Physics, University of Washington, Seattle, WA 98195, USA}
\alsoaffiliation[co-first]
{Authors contributed equally to this work}

\author{Can Liao}%epsilon tests, scientific applications
\affiliation[University of Washington]
{Department of Chemistry, University of Washington, Seattle, WA 98195, USA}
\alsoaffiliation[co-first]
{Authors contributed equally to this work}

\author{Shiv Upadhyay}%advised on framework
\affiliation[University of Washington]
{Department of Chemistry, University of Washington, Seattle, WA 98195, USA}
\alsoaffiliation[co-first]
{Authors contributed equally to this work}

\author{Hang Hu}%das developer, assisted with framework dev
\affiliation[University of Washington]
{Molecular Engineering and Sciences Institute, University of Washington, Seattle, WA 98195, USA}

\author{Tianyuan Zhang}%original das paper
\affiliation[University of Washington]
{Department of Chemistry, University of Washington, Seattle, WA 98195, USA}

\author{A. Eugene DePrince III}
\affiliation[FSU]{Department of Chemistry and Biochemistry, Florida State University, Tallahassee, FL 32306, USA}

\author{Chao Yang}% Invaluable advice about Applied Math/Davidson method
\affiliation[LBNL]{Applied Mathematics and Computational Research Division, Lawrence Berkeley National Laboratory, Berkeley, CA 94720, USA}

\author{Xiaosong Li}%PI
\affiliation[University of Washington]
{Department of Chemistry, University of Washington, Seattle, WA 98195, USA}
\email{xsli@uw.edu}
\begin{document}

%-------------------------------------------------------------------------------
% ABSTRACT
%-------------------------------------------------------------------------------
\twocolumn[
\begin{@twocolumnfalse}
\begin{abstract}
The combinatorial growth of configuration interaction (CI) has long limited this formally exact quantum chemistry method to only the smallest molecules. 
Here, we report a numerically exact CI calculation exceeding one quadrillion ($10^{15}$) determinants, made possible by a lossless categorical compression strategy within the small-tensor-product distributed active space (STP-DAS) framework. 
This approach overcomes the traditional memory bottlenecks of CI by a numerically exact compression of the wavefunction representation and reformulating the most computationally demanding matrix--vector operations.  
Using this method, we performed a fully relativistic CI calculation of the ground state of HBrTe with over $10^{15}$ complex-valued determinants in just 34.5 hours on 1000 computing nodes---the largest CI calculation ever reported. 
We further achieved fast computation for systems with hundreds of billions of determinants on only a few compute nodes.  
Extensive benchmarks confirm that the method retains full numerical exactness while cutting memory and computational cost by orders of magnitude. 
Compared to previous state-of-the-art CI calculations, this work achieves a 1,000 times increase in CI space, a 10$^6$-fold increase in floating-point operations performed, and a 10$^6$-fold improvement in computational speed.  

\end{abstract}
\end{@twocolumnfalse}
]

\section{Introduction}

Full configuration interaction (FCI) offers the most complete and accurate description of a molecule’s electronic structure within a given basis set, providing the exact spectral solution to the non-relativistic electronic Schr\"odinger equation.\cite{Shavitt77_book,Shavitt98_3,Schaefer99_143,Carsky02_book,Shavitt03_book,Shepard12_108}
Due to its variational nature, FCI is particularly well-suited for treating relativistic effects, such as spin--orbit and spin--spin couplings beyond perturbation theory,\cite{Faegri96_4083,Jensen08_034109,Fleig08_014108,Fleig10_014108,Reiher14_041101,Shiozaki15_044112,Mizukami15_4733,Shiozaki16_3781,Saue16_074104,Shiozaki18_014106,Knecht18_2353,Saue19_40,Li20_2975,Li22_141,Li22_2947,Li22_2983,Li23_044101} which are fundamentally rooted in the electronic Dirac equation.

At its core, solving the FCI problem reduces to diagonalizing a large many-electron Hamiltonian matrix. This matrix is Hermitian, sparse, and typically diagonally dominant, making it well-suited to iterative diagonalization techniques that can efficiently converge on a few eigenstates without requiring full storage or construction of the entire matrix. However, as the number of determinants grows factorially with system size, reflecting the combinatorial nature of Slater determinant enumeration within the full Hilbert space, even iterative methods become intractable beyond a certain threshold.

The CI wavefunction is often expressed as a linear combination of Slater determinants, typically generated by excitations from a mean-field self-consistent field (SCF) ground-state reference. In the relativistic regime, this framework must be reformulated using complex-valued 2- or 4-spinor wavefunctions, as required by the Dirac formalism.\cite{Dyall07_book,Wolf15_book}

\begin{figure}[ht!]
    \centering
    \captionsetup{width=.95\linewidth}
    \includegraphics[width=0.95\linewidth]{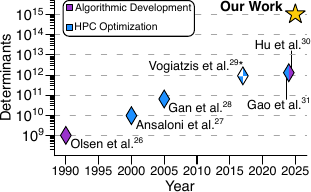}
    \caption{The evolution of the state-of-the-art for CI calculations over time\cite{Simons90_463, Rossi00_496, Harrison05_22, DeJong17_184111,Li24_041404, Yoshida24_1185}. Each historical point is colored according to the nature of the key development of the respective work, which we classify as either intrinsic algorithmic developments (in purple) or optimizations on the HPC hardware of the time (in blue). $^*$ Only one CI iteration was performed. {Source data are provided as a Source Data file.}
}
\label{fig:history}
\end{figure}

\begin{figure*}[ht!]
    \centering
    \includegraphics[width=1\linewidth]{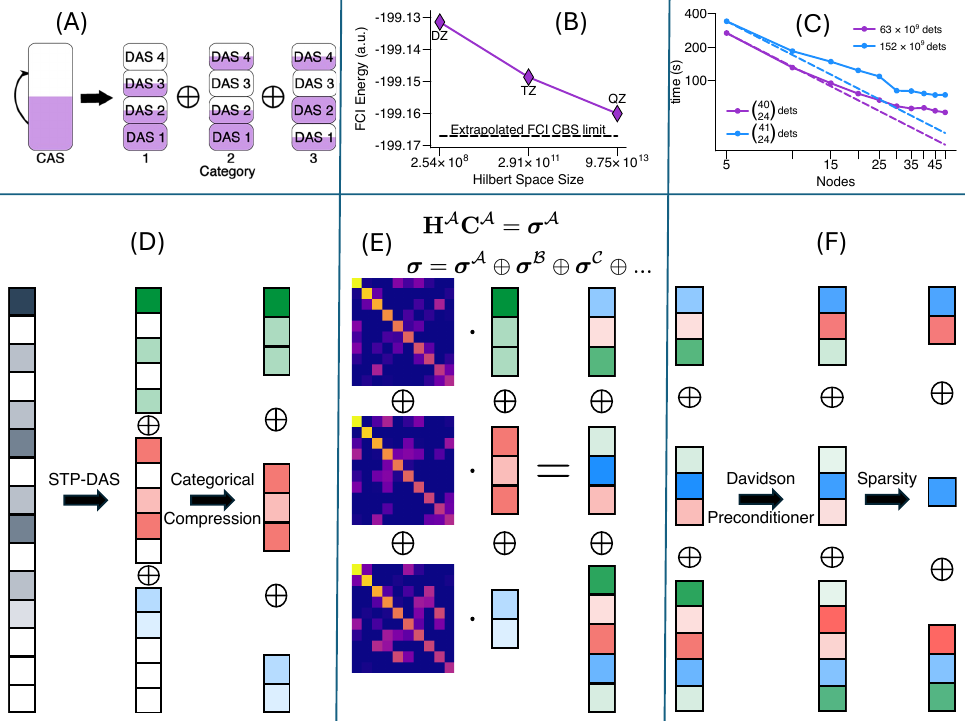}
    \caption{Categorical Compression within the small-tensor-product distributed active space (STP-DAS) framework. (A), The STP-DAS framework decomposition of a complete active space configuration interaction (CASCI) calculation into a direct sum of categorical excitations. The large excitation lists can be factored into much smaller categorical excitation lists. Purple sections within active spaces represent electron-occupied orbitals.
(B), The exact two-component full configuration interaction (X2C-FCI) ground state energy of the Mg\textsuperscript{2+} ion within the cc-pVNZ-DK\cite{Dixon01_48,Wilson10_69}($N=2,3,4$) basis sets, along with the extrapolated complete basis set limit. {Source data are provided as a Source Data file.}
(C), Average execution time (in seconds) of the compression-compatible STP-DAS algorithm per Davidson iteration of a thallium hydride (TlH) test case versus the node count (5 iterations, 1 message passing interface (MPI) process per node, 40 symmetric multiprocessing (SMP) threads per MPI process). {Here, $\mathbf{H}$ is the Hamiltonian, $\mathbf{C}$ is the CI vector, and $\boldsymbol\sigma$ is their product}. The dashed lines illustrate the ideal strong scaling behavior of each CASCI calculation. {Source data are provided as a Source Data file.}
(D), The representation of the subspace expansion vector in a traditional configuration interaction (CI) picture, the decomposed subspace vector in the STP-DAS framework, and the {numerically exact} compression of the subspace expansion vector in the categorically compression-compatible STP-DAS representation. {The color of CI coefficients indicates their configuration category, while their brightness symbolizes their magnitude. White indicates a magnitude of zero.}
(E), A schematic representation of the lossless, compression-compatible, STP-DAS $\boldsymbol\sigma$-build algorithm. {The Hamiltonian matrix is represented as a heatmap, where brighter elements have larger magnitudes. The color of the vector elements indicates the configuration category of the corresponding CI coefficients. Note that the $\boldsymbol\sigma$-build preserves categorical compression.}
(F), An illustration comparing the traditional Davidson preconditioner with the compression-compatible preconditioner to generate successive subspace expansion vectors. The compression-compatible preconditioner appends the subspace with the same effective search direction as the traditional Davidson preconditioner without compromising its compression.
}
\label{fig:SparseDASpanel}
\end{figure*}

\Cref{fig:history} illustrates the historical progression of CI implementations, highlighting major breakthroughs in the achievable scale of determinants. Prior to this work, over a span of 35 years (1990--2024), the field advanced from handling billions to trillions of determinants, driven largely by advances in computer hardware technologies. Although CI is amenable to large-scale parallel processing schemes,\cite{Simons90_463,Rossi00_496,Harrison05_22,DeJong17_184111,Yoshida24_1185, Li24_041404} the explosive growth in memory requirements has historically restricted its applicability to only the smallest chemical systems. Relativistic CI is even more limited, due to the intrinsically larger spinor configuration space associated with complex-valued 2- or 4-component wavefunctions.  Simply put, enabling CI for practical quantum chemistry applications demands {alternative} theoretical frameworks and data representations that can circumvent the brute-force enumeration of the CI space.

%%approximate methods exist to overcome this cost
Many CI-based wavefunction methods aim to approximate the FCI solution. These methods use different types of approximations, which affect the accuracy of the resulting wavefunction. The two main approaches are complete active space CI (CASCI) and selected CI (SCI). 
While both CASCI and SCI methods effectively truncate the Hilbert space of the system to a subspace of significant determinants, this significance is determined differently and at different stages of the computation. 

In the CASCI method,\cite{Handy84_315,Roos87_399,Jensen88_2185,Celani00_5653,Marian01_4775,Visscher03_2963,Visscher06_104106,Fleig08_014108,Fleig10_014108,DeJong17_184111,Yoshida24_1185} it is assumed that only a subspace of the full Hilbert space of the system contains meaningful correlation, and the FCI wavefunction is approximated as the CI wavefunction in the truncated space (the so-called active space). 
The truncation often leads to an underestimation of dynamic correlation.
{Applying more computationally demanding methods such as multiconfigurational self-consistent field (MCSCF)\cite{Faegri96_4083, Frisch03_713, Gagliardi11_044128, Gagliardi15_3010, Jensen08_034109, LiManni22_251, Malmqvist86_479, Meyer81_5794, Mizukami15_4733, Neese13_104113, Roos89_189, Schaefer18_1235, Shepard11_191, Shepard12_108, Shiozaki15_044112, Shiozaki18_014106}, multireference configuration interaction (MRCI)\cite{Liu73_1925,Sigbahn80_157,Shavitt81_91,Knowles88_5803,Roos90_5477,Hirao99_book,Shepard11_191,Shepard12_108,Li20_2975}, and many-body perturbation theory (MRPT2, CASPT2, NEVPT2, MC-PDFT)\cite{Wolinski90_5483,Roos92_1218,Merchan_96_219,Werner00_5546,Malrieu01_10252,Mizukami15_4733,Gagliardi16_3208,Shiozaki16_3781,Li22_2983,gagliardi14_3669} is typically required to achieve qualitative and quantitative agreement with experiment.}

SCI-based methods estimate the importance of each configuration in the total wavefunction based on a predefined significance criterion, which depends on the chemical problem of interest.\cite{Ernest69_23,Hackmeyer69_5584,Rancurel73_5745,Malrieu83_91,Ruedenberg09_64,Umrigar16_3674,Evangelista16_161106,Whaley16_044112,Hoffmann16_1169,Umrigar17_1595,Sharma17_164111,Ayers18_66,Umrigar18_214110,Timothy19_4834,Hoffmann20_2296,Li21_5482,Hoffmann21_949}
Once the most significant determinants are identified, the Hamiltonian is constructed and diagonalized within this reduced space to approximate the FCI wavefunction. As in CASCI, SCI approaches are often combined with perturbation theory to recover contributions from neglected configurations and improve quantitative accuracy.\cite{Rancurel73_5745,Malrieu83_91,Umrigar16_3674,Whaley16_044112,Evangelista16_161106}

Even with advances in dimensionality reduction, CI remains fundamentally constrained by memory limitations. State-of-the-art implementations still require explicit storage of either Hamiltonian matrix elements or excitation lists to support on-the-fly matrix–vector operations, commonly referred to as the $\boldsymbol\sigma$-build.\cite{Handy84_315, Sieghban84_417, Olsen88_2185, Zarrabian89_183} For large CI spaces, storing the full Hamiltonian matrix and performing direct diagonalization is clearly impractical.
In on-the-fly CI algorithms, the one-electron excitation list, which encodes the allowed excitations between determinants for efficient Hamiltonian construction, scales as $n_e \times (n_h + 1) \times N$, where $N$ is the number of determinants, and $n_e$ and $n_h$ denote the numbers of electrons and holes (unoccupied orbitals), respectively. Since $N$ increases factorially with system size, the associated memory requirements grow rapidly, making conventional CI calculations infeasible for anything beyond the smallest systems.

For a CI problem involving $N$ determinants, the size of each CI expansion vector scales linearly with $N$. For instance, a relativistic CI problem with one quadrillion ($10^{15}$, 100 orbitals, 88 electrons) determinants would require approximately 16 petabytes (PB) of memory just to store a single CI vector composed of complex-valued double-precision coefficients. 
While this memory footprint alone makes such problems challenging to tackle, the memory required to store the excitation list can easily scale to the exabyte (EB) regime. This poses a fundamental barrier to scalability, even before considering computational cost. 

A recently introduced CI matrix–vector product algorithm\cite{Li24_041404} leverages the exact factorization of the active space into small tensor products of distributed active spaces---an approach known as the small-tensor product distributed active space (STP-DAS) framework, illustrated in \Cref{fig:SparseDASpanel}(A). {The STP-DAS algorithm reformulates the large CI matrix–vector product into a sequence of small tensor products, each embedded within a distributed active space, computed on-the-fly using string-based methods. This advance exploits the mathematical condition governing the phase relationship between the global address and the local DAS address of any CI matrix element, enabling the use of only small local determinant address strings in the CI matrix–vector build and overcoming the memory bottleneck associated with storing the full excitation list.} This formulation enables extensive reuse of Hamiltonian excitation lists, leading to a dramatic reduction in memory demands. For a CI problem involving one quadrillion ($10^{15}$) determinants, the STP-DAS framework reduces the excitation list memory requirement from 12 exabytes (EB) to 25 gigabytes (GB), an 8-orders-of-magnitude reduction! {In addition, by evenly distributing the computation of small tensor products, the STP-DAS algorithm achieves excellent load balance with minimal node-to-node communication overhead, ensuring strong scalability across both single-node and large-scale parallel architectures.}

\begin{table*}[!ht]
%    \footnotesize
    \caption{Details of the convergence of the X2C-CASCI calculation (100 2-spinor orbitals, 88 electrons, $10^{15}$ 2-spinor determinants) of the ground state of HBrTe (x2c-TZVPall\cite{Weigend17_3696}). Each row corresponds to a Davidson iteration. The fourth and fifth columns track the changes in the computed Rayleigh–Ritz eigenpair across iterations, and the sixth column tracks the residual norm, $\|\mathbf{r}\|=\|\mathbf{HC}-E\mathbf{C}\|$, where $\mathbf{H}$ is the Hamiltonian, $\mathbf{C}$ is the CI vector, $\boldsymbol\sigma$ is their product, and $E$ is the corresponding energy. The preconditioning dropping threshold was taken to be $\varepsilon=1.0\times10^{-5}$.}
    \label{tab:HBrTeCASCI}
% \resizebox{\columnwidth}{!}{
\begin{tabular}{cclccc}
\toprule
Iteration&Duration (s)&Energy (E\textsubscript{h})&$\Delta E$ (E\textsubscript{h})&$\text{max}\{|\Delta C_i|\}$& $\|\mathbf{r}\|$ (E\textsubscript{h})\\ 
\hline
1&122&$-${9812}.080102321897&$-$&$-$&$2.62\times10^{-1}$\\
2&111&$-${9812.1}29587025540&$4.95\times10^{-2}$&$5.45\times10^{-2}$&$8.73\times10^{-2}$ \\
3&262&$-${9812.133}341658482&$3.75\times10^{-3}$&$1.36\times10^{-2}$&$3.23\times10^{-2}$ \\
4&710&$-${9812.1337}35698319&$3.94\times10^{-4}$&$2.26\times10^{-3}$&$1.40\times10^{-2}$ \\
5&1536&$-${9812.1337}84486688&$4.88\times10^{-5}$&$1.08\times10^{-3}$&$7.14\times10^{-3}$ \\
6&3293&$-${9812.13379}0409140&$5.92\times10^{-6}$&$4.97\times10^{-4}$&$4.09\times10^{-3}$ \\
7&6432&$-${9812.133791}245860&$8.37\times10^{-7}$&$1.31\times10^{-4}$&$2.54\times10^{-3}$ \\
8&11565&$-${9812.1337914}08300&$1.62\times10^{-7}$&$9.17\times10^{-5}$&$1.61\times10^{-3}$ \\
9&20384&$-${9812.13379145}1150&$4.28\times10^{-8}$&$4.25\times10^{-5}$&$9.71\times10^{-4}$ \\
\bottomrule
\end{tabular}
% }
\end{table*}

Since the excitation list typically dominates the storage requirements in CI calculations, the STP-DAS framework overcomes a longstanding memory bottleneck and enables CI computations that were previously deemed intractable. By effectively eliminating the memory footprint of the excitation lists, the storage of CI vector coefficients now emerges as the bottleneck in many large-scale CI problems.

Revisiting the earlier CI example of $10^{15}$ determinants: even after eliminating the memory bottleneck associated with storing excitation lists, the same calculation still demands 16~PB of memory to store the numerically exact coefficients of a single CI expansion vector in an iterative solver. Given that practical CI calculations typically require multiple subspace vectors for convergence, it becomes clear that storing these expansion vectors now represents the dominant memory bottleneck in large CI calculations. To address this challenge, we leverage the locally compressed nature of the STP-DAS framework to efficiently compute ultra-large-scale CI problems involving up to a quadrillion ($10^{15}$) determinants. This approach yields deterministic, numerically exact solutions and effectively shifts CI calculations from being memory-bound to compute-bound.

\section{Results}

{
Before presenting methodological details and performance benchmarks, we highlight the largest CASCI calculation to date for the ground state of HBrTe, enabled by the compression-compatible STP-DAS framework to be introduced herein. HBrTe is a substituted form of a hydrogen chalcogenide where one of the hydrogens was substituted with a bromine atom to decrease the symmetry to the molecule. We performed a relativistic exact two-component\cite{Dyall97_9618,Dyall98_4201,Liu05_241102,Peng06_044102,Cheng07_104106,Saue07_064102,Peng09_031104,Reiher13_184105,Repisky16_5823,Li17_2591,Cheng21_1536,Li22_2947,Li22_2983,Li22_5011} CASCI (X2C-CASCI) calculation (100 2-spinor orbitals, 88 electrons, complex-valued $1.05\times 10^{15}$ 2-spinor determinants) of the ground state of HBrTe using the compression-compatible STP-DAS framework. We also performed a calculation with the same number of determinants for the ground state of a magnesium atom (see Section S3 of the Supplementary Information). The calculation ran on the National Energy Research Scientific Computing Center's Perlmutter high-performance supercomputer with a total of 1000 nodes (AMD EPYC 7763 Milan, 128,000 compute cores, 512 GB of RAM per node, 200 GB/s NIC, 2 message passing interface (MPI) processes per node, and 64 symmetric multiprocessing (SMP) threads per MPI process).}

The excitation list is generated without assuming any symmetry of the target state. Consequently, the calculation is formally performed in a quadrillion-determinant space, with all determinants explicitly included. Because the CI coefficients are complex, the memory footprint is twice that of an analogous nonrelativistic calculation. Moreover, the complex arithmetic makes the computational cost (in FLOP count) equivalent to that of a nonrelativistic calculation with more than twice as many determinants.

{\Cref{tab:HBrTeCASCI} summarizes the results of the HBrTe calculation. The ground-state energy converged in 9 iterations to microhartree precision ($<10^{-6}$ a.u.) with a total runtime of 34.5 hours. In each iteration, an additional CI expansion vector was introduced to accelerate convergence, while the compression algorithm dynamically adapted to the expanding vector space, leading to a gradual increase in computational cost. A total of 9 expansion vectors were involved in the $\boldsymbol\sigma$-build. Despite the enormous configuration space of over one quadrillion ($1.05\times10^{15}$) complex-valued 2-spinor determinants, the average $\boldsymbol\sigma$-build time per vector remained just 3.8 hours.}

{The ground-state energy of the HBrTe molecule from this CI calculation is $-$9395.0280045 a.u. Leveraging the gap theorem,\cite{DavisKahan70_1,Parlett98_book} we determine that our $\binom{100}{88}$ X2C-CASCI result lies within $10 \times 10^{-6}$ a.u. of the true x2c-TZVPall\cite{Weigend17_3696} ground state energy within that active space, well below any chemically meaningful threshold.} A detailed analysis is provided in Methods.

This work represents a 3-orders-of-magnitude increase in CI space and a 6-orders-of-magnitude increase in FLOP count, which is estimated using $\mathcal{O}(N^2)$, compared to the previous state-of-the-art in CI calculations.\cite{Li24_041404,Yoshida24_1185} Compared to previous state-of-the-art CI calculations, this work also achieves a 6-orders-of-magnitude speedup in time-to-completion, as measured in core seconds per exaFLOP ($\frac{\text{core} \cdot \text{second}}{\text{exaFLOP}}$, see Section S2 in the Supplementary Information for analysis).
{This ultra-large-scale CI calculation is enabled by the STP-DAS-based numerically exact categorical compression scheme, which reduces the memory required to store the 9 CI expansion vectors from 134 PB to less than 500 TB while maintaining a good load balance,\cite{Li24_041404} making the computation feasible on most existing supercomputing infrastructures.}

We now describe the algorithmic developments that enable such CI calculations to be performed on existing supercomputing resources. Detailed algorithms, parallel implementation strategies, and error bound analyses are provided in the Supplementary Information. {The central concept of the STP-DAS framework is the systematic partitioning of the full CI orbital space into a collection of distributed active spaces. Within each active space, configurations are further classified into categorical subspaces, rigorously defined by distinct electron occupation patterns, as illustrated in \Cref{fig:SparseDASpanel}(A)}.\cite{Li24_041404} The STP-DAS framework reformulates the CI $\boldsymbol\sigma$-build as a sum of small-tensor products, each uniquely addressed via a global tensor looping structure. The major memory bottleneck associated with storing the excitation list is eliminated by allowing categorical subspaces to share compact, local excitation lists.

{In the largest CASCI calculation ($10^{15}$ determinants) presented here, employing 13 distributed active spaces, the STP-DAS approach reduces the excitation list memory requirement from $12\times10^{9}$ GB to just 25 GB.
However, storing all nine CI expansion vectors for a system with $10^{15}$ determinants would require approximately 134 PB of memory. On a high-performance computing system such as Perlmutter, this translates to more than 275,000 nodes, each equipped with 512 GB of memory. A straightforward element-wise sparsity treatment, however, does not meet the STP-DAS condition. To overcome this limitation, the following section introduces a categorical compression scheme that achieves the necessary memory reduction, making it possible to carry out large relativistic CI calculations on a medium-sized computing cluster.}

{
With the STP-DAS framework, the memory bottleneck associated with storing CI expansion vectors is effectively eliminated by applying numerically exact categorical compression. The STP-DAS CI expansion vectors take the form
\begin{equation}
    \mathbf{C}=\bigoplus\limits_{\mathcal{B}}\mathbf{C}^\mathcal{B},
\end{equation}
where $\mathcal{B}$ is a category, defined by a unique electron occupation pattern within the distributed active spaces\cite{Li24_041404}. 

The compression scheme, categorical compression (see \Cref{fig:SparseDASpanel}(D)), stores the $\mathbf{C}^\mathcal{B}$ vectors as compressed sparse column (CSC) vectors. In this format, each categorical expansion vector $\mathbf{C}^\mathcal{B}$ is represented by two equally sized arrays: one, $V_\text{CSC}^{\mathbf{C}^\mathcal{B}} \equiv \{C_{K^\mathcal{B}} : C_{K^\mathcal{B}} \neq 0\}$, stores the values of the nonzero coefficients, while the other, $A_\text{CSC}^{\mathbf{C}^\mathcal{B}} \equiv \{K^\mathcal{B} : C_{K^\mathcal{B}} \neq 0\}$, stores their corresponding local addresses. 
The tensor-loop structure in the STP-DAS $\boldsymbol\sigma$-build algorithm is reformulated in terms of categorically compressed local addresses together with their corresponding global phase factors (see \Cref{fig:SparseDASpanel}(E)). 

In contrast to element-wise compression, categorical compression can eliminate all configurations within a category, i.e., skip an entire category at once. This approach is better able to preserve the vectorized structure compared to element-wise compression. The major difficulty of any sparse matrix-vector product algorithm is the lack of a priori knowledge of the location of the nonzero elements. This difficulty leads to a major bottleneck rooted in the nonuniform accesses to memory. The categorical compression localizes nonuniform memory accesses within a category, which is always orders-of-magnitudes smaller than the size of the full CI space. This localized memory access pattern is the central strength of the categorical compression scheme.

Within each category, element-wise compression is still applicable to maximize sparsity. Most importantly, a category-based compression scheme naturally supports distributed small tensor products, taking advantage of the reduced memory footprint and improved parallel load balance of the STP-DAS framework.

In summary, the numerically exact categorical compression introduced here allows the STP-DAS $\boldsymbol\sigma$-build algorithm to bypass entire categories of determinants while preserving both the vectorized structure and the local addressing scheme of STP-DAS. Because the compression is fully lossless, the omitted determinants have no impact on the resulting Ritz eigenvalue--eigenvector pair.
}

The final hurdle in reducing CI memory demands lies in the iterative solver. In CI, the Hamiltonian operator is diagonalized iteratively within the full Hilbert space of the system. As a result, the error in the computed Ritz value directly reflects the missing correlation energy in the associated approximate wavefunction. This enables the Ritz residual, defined as $\mathbf{r} \equiv  \|\mathbf{HC}-E\mathbf{C}\|$, to be computed and appended as an additional CI expansion vector. The norm of the residual provides rigorous bounds on the missing correlation energy relative to the true eigenpair (eigenvector and eigenvalue) of the Hamiltonian in the chosen basis.\cite{DavisKahan70_1,KNOWLES89_513,Handy89_75,Mitrushenkov94_559,Parlett98_book,Minkoff05_90,Surjan08_144101,Ritz09_1,Cotton22_224105} A widely used approach that leverages this principle is the Davidson iterative solver.\cite{Davidson75_87} Other related methods that use residual norm as the convergence criterion include the locally optimal block preconditioned conjugate gradient (LOBPCG) method~\cite{Knyazev01_517}, the Jacobi-Davidson~\cite{VanDerVorst96_401} method and generalized preconditioned locally harmonic residual (GPLHR) method~\cite{Xue16_A500,Li18_2034}.

By applying numerical or convergence thresholds at various stages of the Davidson method,\cite{Davidson75_87} one can exploit the sparsity of newly generated CI expansion vectors. With sufficiently tight thresholds, these vectors can span the part of the Hilbert space required to accurately represent the desired wavefunction and drive the iterative diagonalization to any desired level of precision. 
The Davidson method\cite{Davidson75_87} utilized the Davidson preconditioner, which generates the $i$th component of the next trial expansion vector according to
\begin{align}\label{eq:davidsonPreconditioner}
t_i\leftarrow
\begin{cases}
    0, & \text{if } |\lambda-H_{ii}| \text{ is small}\\
    \frac{r_i}{\lambda-H_{ii}}, & \text{else}
\end{cases}
\end{align}
Here, $\lambda$ is the Ritz value of the current iteration, $H_{ii}$ is the $i$th element of the diagonal of the Hamiltonian $\mathbf{H}$, and $r_i$ is the $i$th component of the current residual. Among the various preconditioners employed in the Davidson method,\cite{KNOWLES89_513,Handy89_75,Li11_3540,Li15_4146,Cotton22_224105,Parker24_6738} compression-compatible preconditioners, which discard terms $t_i$ below a numerical threshold $\varepsilon$, have been shown to achieve convergence to the exact same results as the traditional Davidson preconditioner.\cite{KNOWLES89_513,Handy89_75,Mitrushenkov94_559,Surjan08_144101,Yang21_e2341} 

In this work, we apply the compression-compatible categorical preconditioner
\begin{align}\label{eq:sparsePreconditioner}
s_i&\leftarrow
\begin{cases}
    \frac{r_i}{\lambda-H_{ii}}, & \text{if } |\lambda-H_{ii}| \geq 10^{-12}\\
    0, & \text{else}
\end{cases}
\nonumber \\%\implies
t_i&\leftarrow
\begin{cases}
    s_i, & \text{if } |s_i| \geq \varepsilon \|\mathbf{s}\|\\
    0, & \text{else}
\end{cases}
\end{align}
using the nonzero residual elements $r_i$. We do this on-the-fly to avoid explicitly storing the prohibitively large diagonal of $\mathbf{H}$ (a dense vector of size $N_\text{dets}$). \Cref{fig:SparseDASpanel}(F) illustrates the expansion of the CI vector space enabled by the compression-compatible categorical preconditioner used in the Davidson method implemented here.%hereblahPreCond
\Cref{eq:sparsePreconditioner} closely resembles the preconditioner proposed in Ref.~\citenum{Yang21_e2341}, with the key distinction being the inclusion of the Davidson-preconditioned residual norm $\mathbf{\|s\|}$ in the dropping criterion. This facilitates dynamic threshold adjustment: as the iterations progress and $\mathbf{\|s\|}$ decreases, the criterion becomes more stringent. More importantly, the factor of $\mathbf{\|s\|}$ ensures that numerical thresholding is applied to the generated expansion vectors relative to the total norm of their exact (traditional Davidson) counterparts, rather than some fixed cutoff on the absolute values of their entries. This results in a very accurate CI space expansion scheme at the cost of computing and contracting a significant number of Hamiltonian matrix elements (to evaluate $s$ exactly), which would be intractable without the STP-DAS framework.

Algorithms and pseudocodes of the compression-compatible STP-DAS method are presented in Methods, along with discussions on load balancing and parallel implementation. The convergence behavior of the STP-DAS framework equipped with the compression-compatible preconditioner defined in \Cref{eq:sparsePreconditioner} was evaluated across three systems with varying degrees of electron correlation: the magnesium atom, diatomic nitrogen, and a model carbon nanotube. The results are provided in Section S1 of the Supplementary Information.

The analysis reveals that overly aggressive thresholding can cause the Davidson procedure to stagnate, thereby preventing convergence to the correct electronic wavefunction. When thresholds are too loose, newly generated trial vectors quickly become linearly dependent on the existing subspace vectors, signaling that the span of the modified subspace has saturated before achieving convergence. 
However, when the preconditioning threshold satisfies
\begin{equation}
\varepsilon \lessapprox \frac{10}{\sqrt{N_{\mathrm{dets}}}},
\end{equation}
the resulting energies agree with their exact values to better than $10^{-7}$ E$_h$, and the residual norms become correspondingly small, demonstrating successful and reliable convergence.

The CI wavefunction of highly correlated systems is comprised of a large number of determinants with small CI coefficients. Because the Hilbert space of the problem is never truncated, and no determinants are discarded from the wavefunction itself. As a result, the compression-compatible preconditioner easily facilitates convergence to the exact wavefunction, provided that the preconditioning threshold $\varepsilon$ is sufficiently small for the true eigenvector to be accurately represented in the subspace spanned by the modified expansion vectors. 

With the capability to perform large CI calculations, energetic extrapolation to the correlation limit becomes feasible for many-electron systems. \Cref{fig:SparseDASpanel}(B) illustrates the correlation-consistent extrapolation for the Mg${}^{2+}$ ion using double-zeta (DZ, 36 orbitals), triple-zeta (TZ, 68 orbitals), and quadruple-zeta (QZ, 118 orbitals) basis sets, involving $2.54\times 10^8$, $2.91\times 10^{11}$, and $9.75\times 10^{13}$ 2-spinor determinants, respectively. The complete basis set limit of $-$199.16704295~a.u. was obtained using a mixed Gaussian extrapolation scheme tailored for correlation-consistent basis sets.\cite{Dunning94_7410}

To demonstrate the strong scaling behavior of the compression-compatible STP-DAS $\boldsymbol\sigma$-build algorithm, we performed relativistic X2C-CASCI calculations on the thallium hydride (TlH) molecule using active spaces of 40 and 41 2-spinor orbitals and 24 electrons, $\binom{40}{24}$ and $\binom{41}{24}$, corresponding to 63 and 152 billion 2-spinor determinants, respectively. Five distributed active spaces (DASs) were employed. These calculations were executed on the University of Washington’s Hyak HPC system, a small-sized cluster where each node is equipped with two Intel Xeon 6230 Gold CPUs and a single 100 GB/s network interface card. As shown in \Cref{fig:SparseDASpanel}(C), even with just 5 compute nodes, relativistic X2C-CASCI calculations involving tens to hundreds of billions of determinants require only 4-6 minutes per $\boldsymbol{\sigma}$-build on average. Increasing to 30 nodes further reduces the cost to just over one minute. Past 30 nodes, the communication time dominates the runtime and the calculations no longer scale. This benchmark demonstrates that billion- and even trillion-determinant CI calculations are now feasible on a small-scale computing cluster. 

{
Additionally, we performed X2C-CASCI calculations for the ground states of two highly correlated systems, showcasing the applicability of the compression-compatible STP-DAS framework to highly correlated systems. These systems were chosen from opposite ends of the correlation spectrum from strongly statically correlated to strongly dynamically correlated. Square Rb$_4$, a relativistic analogue of H$_4$\cite{Stevens73_3378,Clary96_8413,Hellgren24_074106,Noe21_124108,Nakatsuji23_140359,Champagne12_024315,Noga11_418,Fromager22_032203,Ellis11_124108,Head-Gordon00_8873,Sorella19_084102,Whaley23_030307,Mazziotti13_44}, displays strong static correlation and Xe$_2$ is a dynamically correlated noble gas dimer\cite{Saue12_54,Hobza96_425,Yang97_7921,Evangelisti03_303,Malijevsky03_2102,Perdew05_114102,Stoll05_3917,Becke09_719,Angyan10_244108}.

\Cref{tab:Rb$_4$,tab:Xe$_2$} summarize the results of the Rb$_4$ and Xe$_2$ calculations. The Rb$_4$ calculation (50 2-spinor orbitals, 28 electrons, $8.9\times 10^{13}$ 2-spinor determinants) ran on the National Energy Research Scientific Computing Center's Perlmutter high-performance supercomputer with a total of 100 nodes (AMD EPYC 7763 Milan, 12,800 compute cores, 512 GB of RAM per node, 200 GB/s NIC, 2 MPI processes per node, and 64 SMP threads per MPI process) and took 6 iterations and 11.8 hours to converge. The Xe$_2$ calculation (60 2-spinor orbitals, 12 electrons, $1.4\times 10^{12}$ 2-spinor determinants) ran on the same platform with 256 nodes and took 7 iterations and 36.1 hours to converge.

The ground-state energies of the Rb$_4$ and Xe$_2$ molecules from these calculations are $-$11916.1527249 a.u. and $-$14889.6466956 a.u., accordingly. The gap theorem\cite{DavisKahan70_1,Parlett98_book} guarantees that these X2C-CASCI results lie within $0.53 \times 10^{-6}$ a.u. and $17.56 \times 10^{-6}$ a.u. of the true ground state energies within the corresponding active spaces and basis sets (see Methods).
}
\begin{table*}[!ht]
%    \footnotesize
    \caption{Details of the convergence of the X2C-CASCI calculation (50 2-spinor orbitals, 28 electrons, $8.9\times10^{13}$ 2-spinor determinants) of the ground state of Rb$_4$ (cc-pvtz-x2c\cite{Peterson17_244106}). Each row corresponds to a Davidson iteration. The third and fourth columns track the changes in the computed Rayleigh–Ritz eigenpair across iterations, and the fifth column tracks the residual norm, $\|\mathbf{r}\|=\|\mathbf{HC}-E\mathbf{C}\|$, where $\mathbf{H}$ is the Hamiltonian, $\mathbf{C}$ is the CI vector, $\boldsymbol\sigma$ is their product, and $E$ is the corresponding energy. The preconditioning dropping threshold was taken to be $\varepsilon=1.0\times10^{-6}$.}
    \label{tab:Rb$_4$}
% \resizebox{\columnwidth}{!}{
\begin{tabular}{clccc}
\toprule
Iteration&Energy (E\textsubscript{h})&$\Delta E$ (E\textsubscript{h})&$\text{max}\{|\Delta C_i|\}$& $\|\mathbf{r}\|$ (E\textsubscript{h})\\ 
\hline
1&$-${90.0}38089795466&$-$&$-$&$4.48\times10^{-2}$\\
2&$-${90.0}48919589524&$1.08\times10^{-2}$&$1.21\times10^{-1}$&$1.74\times10^{-2}$ \\
3&$-${90.0}49940164869&$1.02\times10^{-3}$&$1.64\times10^{-2}$&$4.95\times10^{-3}$ \\
4&$-${90.05000}2638482&$6.25\times10^{-5}$&$2.53\times10^{-3}$&$1.20\times10^{-3}$ \\
5&$-${90.050006}376495&$3.74\times10^{-6}$&$1.37\times10^{-3}$&$2.96\times10^{-4}$ \\
6&$-${90.0500065}93144&$2.17\times10^{-7}$&$1.55\times10^{-4}$&$7.36\times10^{-5}$ \\
\bottomrule
\end{tabular}
% }
\end{table*}

\begin{table*}[!ht]
%    \footnotesize
    \caption{Details of the convergence of the X2C-CASCI calculation (60 2-spinor orbitals, 12 electrons, $1.4\times10^{12}$ 2-spinor determinants) of the ground state of Xe$_2$ (x2c-TZVPall-2c\cite{Weigend17_3696}). Each row corresponds to a Davidson iteration. The third and fourth columns track the changes in the computed Rayleigh–Ritz eigenpair across iterations, and the fifth column tracks the residual norm, $\|\mathbf{r}\|=\|\mathbf{HC}-E\mathbf{C}\|$, where $\mathbf{H}$ is the Hamiltonian, $\mathbf{C}$ is the CI vector, $\boldsymbol\sigma$ is their product, and $E$ is the corresponding energy. The preconditioning dropping threshold was taken to be $\varepsilon=1.0\times10^{-5}$.}
    \label{tab:Xe$_2$}
% \resizebox{\columnwidth}{!}{
\begin{tabular}{clccc}
\toprule
Iteration&Energy (E\textsubscript{h})&$\Delta E$ (E\textsubscript{h})&$\text{max}\{|\Delta C_i|\}$& $\|\mathbf{r}\|$ (E\textsubscript{h})\\ 
\hline
1&$-${21}.059579374232&$-$&$-$&$4.76\times10^{-1}$\\
2&$-${21}.194173478231&$1.35\times10^{-1}$&$3.90\times10^{-2}$&$1.86\times10^{-1}$ \\
3&$-${21.20}6728805229&$1.26\times10^{-2}$&$8.89\times10^{-3}$&$5.62\times10^{-2}$ \\
4&$-${21.207}569776081&$8.41\times10^{-4}$&$8.74\times10^{-4}$&$1.75\times10^{-2}$ \\
5&$-${21.2076}34517148&$6.47\times10^{-5}$&$4.51\times10^{-4}$&$7.88\times10^{-3}$ \\
6&$-${21.20764}4283591&$9.77\times10^{-6}$&$1.19\times10^{-4}$&$4.48\times10^{-3}$ \\
7&$-${21.207646}647756&$2.36\times10^{-6}$&$3.90\times10^{-5}$&$2.93\times10^{-3}$ \\
\bottomrule
\end{tabular}
% }
\end{table*}

In summary, by combining compression-compatible preconditioners with compression-compatible categorical CI vectors, the STP-DAS framework drastically reduces the memory footprint of both the excitation lists and the CI expansion vectors. {In the largest relativistic CASCI calculation presented here, spanning $10^{15}$ determinants across 13 distributed active spaces, the STP-DAS approach reduces the memory required for the excitation list from $12 \times 10^9$ GB to just 25 GB, and for 9 CI expansion vectors from 134 PB to less than 500 TB. These reductions make quadrillion-determinant calculations tractable on current supercomputing architectures. While most of the community may not have access to the hundreds of compute nodes required for such runs, this work also demonstrates the practical feasibility of trillion-determinant calculations on just a few nodes and even on a laptop.}

\section{Discussion}

In this work, we conducted a relativistic configuration interaction (CI) calculation for the ground state of HBrTe in a quadrillion-determinantal space. This calculation was enabled by {numerically exact} categorical compression within the STP-DAS framework, which effectively eliminates the memory bottlenecks associated with storing both excitation lists and CI expansion vectors. Compared to previous state-of-the-art CI calculations, this work represents a 3-orders-of-magnitude increase in CI space, a 6-orders-of-magnitude increase in FLOP count, and a 6-orders-of-magnitude improvement in computational speed.

We introduced a categorically compressed representation of the CI expansion vectors and reformulated the STP-DAS $\boldsymbol{\sigma}$-build algorithm to take advantage of this structure. By expressing the global expansion vector as a direct sum of compressed local components, the algorithm efficiently skips all coefficients that do not contribute to the categorical $\boldsymbol{\sigma}$-vector. This approach is further enabled by a compression-compatible preconditioner, which generates compressed expansion directions within the Davidson procedure.

The resulting categorically compressed STP-DAS $\boldsymbol{\sigma}$-build algorithm demonstrates excellent strong scaling behavior and yields dramatic reductions in both runtime and memory footprint. These benefits extend seamlessly to both relativistic (two- and four-component) and non-relativistic CI calculations. {To highlight this capability, we computed the $\binom{100}{88}$ X2C-CASCI ground-state energy of HBrTe using over one quadrillion ($10^{15}$) complex-valued 2-spinor determinants. The categorically compressed STP-DAS approach spans $10^{15}$ determinants across 13 distributed active spaces, reducing the memory required for the excitation list from $12 \times 10^9$ GB to only 25 GB, and for nine CI expansion vectors from 134 PB to under 500 TB. It converges the ground-state wave function of HBrTe in just nine iterations over a 34.5-hour runtime. This achievement represents the largest CI calculation reported to date.}
Additionally, we achieved $\boldsymbol{\sigma}$-build times of just 5 minutes for systems with approximately 150 billion complex-valued 2-spinor determinants using only a few compute nodes. The capability to perform large CI calculations makes basis set extrapolations to the complete basis set limit and computations on highly correlated molecular systems readily achievable with CI.

{The integration of categorical compression with STP-DAS marks a paradigm shift in tackling large-scale CI problems. As quantum chemistry continues to push the limits of system complexity, the ability to carry out quadrillion-determinant calculations within tractable resource bounds establishes a powerful foundation for studying highly correlated, multireference, relativistic systems. While access to hundreds of compute nodes for quadrillion-determinant calculations may remain out of reach for most of the community, this work demonstrates the practical feasibility of trillion-determinant calculations on a small cluster.

For transition-metal, rare-earth, and heavy-element complexes, such large-scale CI calculations enable predictive simulations of electronic structure properties (bond order, covalency, polarization, etc.), spectroscopic observables (UV/Vis, X-ray, etc.), and reaction pathways, with the full orbital space consisting of both metal and ligand orbitals, treated on an equal footing.

The ability to simulate a full CI space of 100 orbitals on a classical computer not only challenges current notions of quantum supremacy, but also establishes a robust platform for developing and benchmarking quantum algorithms aimed at achieving chemical accuracy.}

\section{Methods}\label{sec:theory}

\subsection{Lossless $\boldsymbol\sigma$-Build using the Categorical Compression of Small Tensor Products}

The categorical $\boldsymbol\sigma$-build algorithm within STP-DAS\cite{Li24_041404} can be reformulated to exploit the categorical compression of the expansion vectors. The compact nature of the categorical representation enables fast and memory-efficient computation of $\boldsymbol\sigma$-vectors. 
The categorical $\boldsymbol\sigma$-build algorithm implements the evaluation of
\begin{align}
    \sigma_{L^{\mathcal{A}}} &=   {}^\text{1e} \sigma_{L^\mathcal{A}} + {}^\text{2e} \sigma_{L^\mathcal{A}}, \label{eq:sigmaL}\\
    {}^\text{1e} \sigma_{L^\mathcal{A}} & = \sum_{\mathcal{B}} \sum_{\mathbb{K}^\mathcal{B}_\mu\oplus \mathbb{K}^\mathcal{B}_\nu}\sum_{pq} P_{\mu\nu} \delta_{\bar{\mathbb{X}}_{\mu\nu}^\mathcal{A}\bar{\mathbb{X}}_{\mu\nu}^\mathcal{B}} \notag\\& h'_{pq} 
    \bra{\mathbb{L}^\mathcal{A}_\mu\oplus \mathbb{L}^\mathcal{A}_\nu}
        \hat{E}_{pq} \ket{\mathbb{K}^\mathcal{B}_\mu\oplus \mathbb{K}^\mathcal{B}_\nu} C_{K^\mathcal{B}},  
    \label{eq:sigmaL1e}\\ 
{}^\text{2e} \sigma_{L^\mathcal{A}} & = 
 \frac{1}{2}  \sum_{\mathcal{C} \mathcal{B}}\sum_{\mathbb{J}^\mathcal{C}_\mu\oplus \mathbb{J}^\mathcal{C}_\nu}\sum_{\mathbb{J}^\mathcal{C}_\kappa\oplus \mathbb{J}^\mathcal{C}_\lambda}
 \sum_{\mathbb{K}^\mathcal{B}_\kappa\oplus \mathbb{K}^\mathcal{B}_\lambda} \sum_{pqrs}  P_{\mu\nu}P_{\kappa\lambda} \notag
\\&
\delta_{\bar{\mathbb{X}}_{\mu\nu}^\mathcal{A}\bar{\mathbb{X}}_{\mu\nu}^\mathcal{C}}\delta_{\bar{\mathbb{X}}_{\kappa\lambda}^\mathcal{C}\bar{\mathbb{X}}_{\kappa\lambda}^\mathcal{B}}  g_{pqrs} 
 \bra{\mathbb{L}^\mathcal{A}_\mu\oplus \mathbb{L}^\mathcal{A}_\nu}
        \hat{E}_{pq} \ket{\mathbb{J}^\mathcal{C}_\mu\oplus \mathbb{J}^\mathcal{C}_\nu} \notag
 \\& \bra{\mathbb{J}^\mathcal{C}_\kappa\oplus \mathbb{J}^\mathcal{C}_\lambda}
        \hat{E}_{rs} \ket{\mathbb{K}^\mathcal{B}_\kappa\oplus \mathbb{K}^\mathcal{B}_\lambda} C_{K^\mathcal{B}},\label{eq:sigmaL2e}
\end{align}
where $p\in\mathbb{X}_\mu^\mathcal{A},~q\in\mathbb{X}_\nu^\mathcal{C}, r\in\mathbb{X}_\kappa^\mathcal{C},~s\in\mathbb{X}_\lambda^\mathcal{B}$, $\bra{\mathbb{X}^\mathcal{A}_\mu\oplus \mathbb{X}^\mathcal{A}_\nu}\hat{E}_{pq} \ket{\mathbb{X}^\mathcal{B}_\mu\oplus \mathbb{X}^\mathcal{B}_\nu}$ are categorical one electron excitation lists, $ P_{\mu\nu}$ are global phase factors, and $h'_{pq}$ ($g_{pqrs}$) are one (two) body Hamiltonian elements. See Ref. \citenum{Li24_041404} for algorithmic details.

\Crefrange{eq:sigmaL}{eq:sigmaL2e} define the categorical $\boldsymbol{\sigma}$-vector in terms of local STP-DAS one-electron excitation lists and the categorical CI expansion vector. Notably, when the expansion coefficients $C_{K^\mathcal{B}}$ are categorically compressed, the resulting $\boldsymbol{\sigma}$-vector coefficients $\sigma_{L^{\mathcal{A}}}$ are also categorically compressed. In such cases, the categorical $\boldsymbol{\sigma}$-build reduces to a contraction between a categorically compressed expansion vector and categorically compressed STP-DAS one-electron excitation lists. Thus, \Crefrange{eq:sigmaL}{eq:sigmaL2e} yield a categorically compressed $\boldsymbol{\sigma}$-vector, in a manner directly analogous to the compression-preserving behavior of sparse matrix–sparse vector products (SpMSpV). Importantly, this compression preservation is general and independent of the specific storage format used for the categorically compressed representations.

The categorically compressed representation can be implemented in various forms, with the choice of storage format guided primarily by computational efficiency. Since the categorical $\boldsymbol{\sigma}$-build algorithm often involves reading numerous expansion coefficients with increasing local addresses during contraction, it is natural to adopt a compressed sparse column (CSC) format for storing the categorical expansion coefficients.

\subsection{Eigenvalue Bound Analysis}\label{ssec:theoryEigenBound}
We wish to apply the gap theorem \cite{DavisKahan70_1,Parlett98_book,Knyazev13_244} to bound the error in the computed X2C-CASCI ground state energy:
\begin{equation}
|\delta E| \leq \frac{\|\mathbf{r}\|^2}{\gamma_0},
\end{equation}
where $\mathbf{r}$ is the Ritz residual of the computed ground state and the gap $\gamma_0 \equiv E_1-\tilde{E}_0$ is the difference between $E_1$, the (unknown) exact energy of the first excited state, and $\tilde{E}_0$, the computed Ritz value of the ground state. 
Because $\gamma_0$ is unknown, one can estimate its order-of-magnitude using approximate methods or use experimental values to compute a surrogate for the true gap. 
One can also obtain an exact lower bound on the gap by including the posterior error bound of the first excited state~\cite{yosida1995_book,saad2011_book,Knyazev13_244} in the Davidson calculation:\cite{Minkoff05_90}
\begin{equation}\label{eq:gammaLowerBound}
    \gamma_0 = E_1-\tilde{E_0}\geq \left(\tilde{E_1}-\|\mathbf{r_1}\|\right)-\tilde{E}_0 \equiv \gamma^-_0,
\end{equation}
where $\mathbf{r_1}$ is the residual associated with the Ritz value $\tilde{E}_1$. 

Using the gap theorem,\cite{DavisKahan70_1,Parlett98_book,Knyazev13_244} we can place an exact bound on the error in the computed X2C-CASCI ground-state energy. This requires an estimate of the energy gap between the ground and first excited states of the X2C-CASCI Hamiltonian. To obtain this, we performed an X2C-CISD calculation for the two lowest-lying states and determined a gap of approximately 0.095 a.u., in good agreement with experimental values. Based on this estimate, the gap theorem bounds the error in our X2C-CASCI ground-state energy to within 10~microhartree, which is well below any chemically meaningful threshold.

\subsection{Compression-Compatible STP-DAS Algorithm}\label{subsec:theoryAlgorithm}

\Cref{alg:sigmaBuild} shows the categorically compressed STP-DAS algorithm, in which only nonzero elements contribute to the categorical $\boldsymbol\sigma$-vector. Its advantage over the traditional STP $\boldsymbol\sigma$-build algorithm\cite{Li24_041404} is twofold: in the outermost loop, where we skip entire categories whose expansion vector vanishes (see line 2-3), and in the inner loop of line 7, where we only process excitations $\langle\mathbb{J}^\mathcal{C}_\kappa\oplus \mathbb{J}^\mathcal{C}_\lambda|\hat{E}_{rs} |\mathbb{K}^\mathcal{B}_\kappa\oplus
\mathbb{K}^\mathcal{B}_\lambda\rangle$ for which $C_{K^\mathcal{B}}\neq 0$.

\begin{algorithm*}%[H]
\DontPrintSemicolon
\caption{Two-electron $\boldsymbol\sigma$-build using categorical compression. {Bold text represents algorithmic logic and typewritten text represents comments.}}\label{alg:sigmaBuild}
\KwData{the categorically compressed expansion vector $\mathbf{C}=\bigoplus\limits_{\mathcal{B}}\left(V_\text{CSC}^{\mathbf{C}^\mathcal{B}},A_\text{CSC}^{\mathbf{C}^\mathcal{B}}\right)$ and integrals $g_{pqrs}$}
\KwResult{the categorically compressed contraction ${}^\text{2e} \boldsymbol\sigma$}
\SetKwFunction{proc}{\textnormal{exponentialSearch}}
\For{{\normalfont categories $\mathcal{A}, \mathcal{B}, \mathcal{C}$ and DAS indices $\mu,\nu,\kappa,\lambda$ for which $ \delta_{\bar{\mathbb{X}}_{\mu\nu}^\mathcal{A}\bar{\mathbb{X}}_{\mu\nu}^\mathcal{C}}\delta_{\bar{\mathbb{X}}_{\kappa\lambda}^\mathcal{C}\bar{\mathbb{X}}_{\kappa\lambda}^\mathcal{B}}=1$ {\bf parallel}}}
{
\If{\normalfont$|A_\text{CSC}^{\mathbf{C}^\mathcal{B}}|=0$}{continue to next iteration}

\For{$\mathbb{J}_{\mu\nu\kappa\lambda}^\mathcal{C}$}{
\tcp{Obtain the sorted DAS one-electron excitation list of $\mathbb{J}_{\mu\nu\kappa\lambda}^\mathcal{C}$}
$A\leftarrow \{\mathbb{K}^\mathcal{B}_\kappa\oplus \mathbb{K}^\mathcal{B}_\lambda : \exists r\in\mathbb{X}_\kappa^\mathcal{C} ~,\exists s\in\mathbb{X}_\lambda^\mathcal{B} \text{ S.T. } \bra{\mathbb{J}^\mathcal{C}_\kappa\oplus \mathbb{J}^\mathcal{C}_\lambda}
        \hat{E}_{rs} \ket{\mathbb{K}^\mathcal{B}_\kappa\oplus \mathbb{K}^\mathcal{B}_\lambda} \neq 0\}$ \;
\tcp{Intersect $A_\text{CSC}^{\mathbf{C}^\mathcal{B}}$ with $A$ to contract nonzero expansion coefficients}
  $i,j\leftarrow 0$\;
  \While{\normalfont $i < |A| \text{ and } j < |A_\text{CSC}^{\mathbf{C}^\mathcal{B}}|$}{

    \If{\normalfont $A[i]<A_\text{CSC}^{\mathbf{C}^\mathcal{B}}[j]$}{
      \tcp{Advance $i$ to the lowest $i'$ for which $A_{i'}\geq A_\text{CSC}^{\mathbf{C}^\mathcal{B}}[j]$}
      \proc{\normalfont $A,i, A_\text{CSC}^{\mathbf{C}^\mathcal{B}}[j]$}
    }
    \uElseIf{\normalfont $A[i]>A_\text{CSC}^{\mathbf{C}^\mathcal{B}}[j]$}{
    \tcp{Advance $j$ to the lowest $j'$ for which $A_\text{CSC}^{\mathbf{C}^\mathcal{B}}[j']\geq A[i]$}
      \proc{\normalfont $A_\text{CSC}^{\mathbf{C}^\mathcal{B}}, j, A[i]$}
    }
    \Else{\normalfont 
    \tcp{Process the $\boldsymbol\sigma$ contribution at the address $L^\mathcal{A}$ into a thread-safe hash table belonging to category $\mathcal{A}$}
    \For{$p\in\mathbb{X}_\mu^\mathcal{A},~q\in\mathbb{X}_\nu^\mathcal{C}$}
      {
        ${}^\text{2e} \sigma_{L^\mathcal{A}} \leftarrow {}^\text{2e} \sigma_{L^\mathcal{A}} + 
        P_{\mu\nu}P_{\kappa\lambda}g_{pqrs}\bra{\mathbb{L}^\mathcal{A}_\mu\oplus \mathbb{L}^\mathcal{A}_\nu}
        \hat{E}_{pq} \ket{\mathbb{J}^\mathcal{C}_\mu\oplus \mathbb{J}^\mathcal{C}_\nu} $ \\
$\qquad \qquad \qquad \quad \ \times \bra{\mathbb{J}^\mathcal{C}_\kappa\oplus \mathbb{J}^\mathcal{C}_\lambda}
        \hat{E}_{rs} \ket{\mathbb{K}^\mathcal{B}_\kappa\oplus \mathbb{K}^\mathcal{B}_\lambda} V_\text{CSC}^{\mathbf{C}^\mathcal{B}}[j]$ \;
      }
      $i\leftarrow i+1$ \;
      $j\leftarrow j+1$ \;
    }
  }
}
}
% \end{algorithm*}
% \begin{algorithm*}%[H]
% \caption{Exponential search subroutine used during the Categorical two-electron $\boldsymbol\sigma$-contraction using categorical compression}\label{alg:sigmaBuildexpsearch}
% \SetKwProg{myproc}{Subroutine:}{\\\KwResult{\textnormal{An index $i < j < |A|$ such that $A[j] \geq x$ or $j=|A|$ if nonexistent}}}{}
\SetKwProg{myproc}{Subroutine:}{}{}
\myproc{\proc{$A, i, x$}}{ 
\KwResult{\textnormal{An index $i < j < |A|$ such that $A[j] \geq x$ or $j=|A|$ if nonexistent}}
$\delta \leftarrow 1$\;
\While{\normalfont $j < |A|$ and $A[j] < x$ }{
  $j\leftarrow j + \delta$\;
  $\delta \leftarrow 2\delta$ \;
}
$j_\text{min}\leftarrow j-\delta/2$\;
$j_\text{max}\leftarrow \text{min}\{j,|A|-1\}$\;
\tcp{Return the lower bound of $x$ in the range $A[j_\textnormal{min} \ldots j_\textnormal{max}]$}
\Return binarySearch{\tt(}$A[j_\text{min} \ldots j_\text{max}], x${\tt)}
}
\end{algorithm*}

There is a potential workload imbalance associated with \Cref{alg:sigmaBuild}: because the collection of categorical expansion vectors is distributed among computing nodes, the contraction workload of a given node is proportional to the number of nonzero categorical expansion coefficients it has. We alleviated some of the resulting computational delay by implementing passive one-sided MPI communication of categorical expansion vectors using remote memory access (RMA). This allows idle nodes to contract more categorical expansion vectors with their excitation lists without waiting for the corresponding busy nodes to broadcast them. Ultimately, overcoming this load-balancing issue requires dynamically redistributing categorical expansion vectors according to their sparsity, which changes during the iterative diagonalization. 

As illustrated in \Cref{alg:sigmaBuild}, the computational cost, both in memory and runtime, of the compression-compatible STP-DAS $\boldsymbol\sigma$-build procedure increases with the density of the expansion vectors. Therefore, it is essential to maintain maximal compression in these vectors. To achieve this, we replace the traditional Davidson preconditioner with a compression-compatible alternative for generating new trial expansion vectors. This modification alters only the subspace expansion strategy in the Davidson algorithm, while preserving exact treatment of the full determinantal space. As a result, the computed matrix–vector product $\mathbf{HC}$ and the corresponding residual norm $\|\mathbf{r}\| = \|\mathbf{HC} - \lambda\mathbf{C}\|$ remain exact, unlike in selected CI and other truncated approaches, where both the Hamiltonian and CI vectors are explicitly approximated.

The compression-compatible STP-DAS $\boldsymbol\sigma$-build algorithm significantly reduces the overall workload associated with the $\boldsymbol{\sigma}$ build. The reduction in workload can be nonuniform: the contraction workload associated with a determinant $\mathbb{J}_{\mu\nu\kappa\lambda}^\mathcal{C}$ is proportional to the number of local addresses containing both nonzero categorical expansion vectors elements and Hamiltonian matrix elements. To improve the load-balance, we implemented dynamic SMP thread-level parallelism in the outermost loop (line 1 in \Cref{alg:sigmaBuild}) instead of in the loop over determinants $\{\mathbb{J}_{\mu\nu\kappa\lambda}^\mathcal{C}\}$ (line 4 in \Cref{alg:sigmaBuild}). Under dynamic parallelism, some threads execute many light contractions, while others execute fewer heavy contractions, resulting in a more uniform distribution of contraction workload. Such dynamic parallelism is ineffective in the loop over determinants $\{\mathbb{J}_{\mu\nu\kappa\lambda}^\mathcal{C}\}$ due to the small tensor product nature of the STP-DAS framework.

\subsection{Data Availability}
All datasets discussed in this work are included in the Supplementary Information/Source Data files. Source data are provided with this paper. {The atomic coordinates for the HBrTe, Rb$_4$, Xe$_2$, Mg, N$_2$, and carbon nanotube systems are available as Supplementary Data 1.}

\subsection{Code Availability}
The compression-compatible STP-DAS framework is implemented in the development version of the Chronus Quantum software package.\cite{Li20_e1436}. The Chronus Quantum software package is open-source software, distributed under the GNU General Public License v2 (GPLv2). All calculations were performed using executables compiled with GCC version 13.2. 
The STP-DAS implementation is currently accessible within the developers' community. It will be included in the next official Chronus public release, following extensive testing across multiple platforms to ensure robustness, scalability, and reproducibility. In the meantime, we created an MPI-compatible, portable, and containerized version of Chronus Quantum containing the compression-compatible STP-DAS implementation. {This version is freely available, along with the relevant input files, in Figshare\cite{Upadhyay25_DockerImage}.}
%\newpage 

%\pagestyle{empty} % uncomment to remove page numbers from references page only
%\renewcommand{\refname}{Type your custom reference section header here} % uncomment to define header for references section; default is "References"

%\bibliography{
%    references/Journal_Short_Name,
%    references/Li_Group_References, 
%    references/CI,
%    references/sparseDAS,
%    references/mcscf,
%    references/rel_other,
%    references/CI_approx
%} % add .bib files to this list
%\newpage

\providecommand{\latin}[1]{#1}
\makeatletter
\providecommand{\doi}
  {\begingroup\let\do\@makeother\dospecials
  \catcode`\{=1 \catcode`\}=2 \doi@aux}
\providecommand{\doi@aux}[1]{\endgroup\texttt{#1}}
\makeatother
\providecommand*\mcitethebibliography{\thebibliography}
\csname @ifundefined\endcsname{endmcitethebibliography}
  {\let\endmcitethebibliography\endthebibliography}{}

\section{Acknowledgements}
The development of variational relativistic multi-reference methods is supported by the U.S. Department of Energy, Office of Science, Basic Energy Sciences, in the Computational and Theoretical Chemistry program (Grant No. DE-SC0006863 to XL). The development of the Chronus Quantum computational software is supported by the Office of Advanced Cyberinfrastructure, National Science Foundation (Grants No. OAC-2103717 and OAC-2103705). XL, CY, and AED acknowledge the support to develop reduced scaling computational methods from the Scientific Discovery through Advanced Computing (SciDAC) program sponsored by the Offices of Advanced Scientific Computing Research (ASCR) and Basic Energy Sciences (BES) of the U.S. Department of Energy (Grant No. DE-SC0022263).
This project used the resources of the National Energy Research
Scientific Computing Center, a DOE Office of Science User Facility
supported by the Office of Science of the U.S. Department of Energy
under Contract No. DE-AC02-05CH11231 using NERSC awards CSGB-ERCAP 0032454 and DDR-ERCAP 0034806.

\section{Author Contribution Statements}
A. Shayit, S. Updhyay, H. Hu, and X. Li conceived the project and developed the computer code. A. E. DePrince and A. Shayit acquired computing resources. C. Liao, A. Shayit, and  S. Updhyay carried out the benchmark calculations and analyzed the data. X. Li and T. Zhang developed the software infrastructure. A. E. DePrince, C. Yang, and X. Li acquired research funding. X. Li, A. Shayit, S. Updhyay, and  C. Liao wrote the manuscript with input from all authors.  All authors discussed the results and approved the final version of the manuscript.

\section{Competing Interest Statement}
The Authors declare no competing interests.

\end{document}